%% file: shear_cross_kappa.tex
\documentclass[aps, prd,twocolumn,superscriptaddress,nofootinbib,preprintnumbers]{revtex4-1}

\input{journaldef.tex}
\include{notations}

\usepackage{amsmath,amssymb,latexsym,times}
\usepackage{numdef}
\usepackage{slashed}
\usepackage{graphicx}
\usepackage{grffile}
\usepackage[normalem]{ulem}
\usepackage{dcolumn}
\usepackage{xcolor}
\usepackage{booktabs}
\usepackage{amssymb}
\usepackage{slashed}
\usepackage{todonotes}
\usepackage{hyperref}
 \hypersetup{
     colorlinks=true,
     linkcolor=aiiro,
     citecolor = aiiro,      
     urlcolor=aiiro,
     }

\usepackage{newtxtext,newtxmath}

\usepackage[T1]{fontenc}
\usepackage{graphicx}	
\usepackage{amsmath}	
\usepackage{amssymb}	
\usepackage{slashed}
\usepackage[normalem]{ulem}
\usepackage{xcolor}
\usepackage{xspace}
\definecolor{aiiro}{HTML}{105779}

\newcommand\Tstrut{\rule{0pt}{2.6ex}}         

\def\Tref#1{Table~\ref{#1}\xspace}

\def\Cref#1{Chapter~\ref{#1}\xspace}

\newcommand{\Planck}{{\slshape Planck~}}
\newcommand{\Planckc}{{\slshape Planck}}

\begin{document}
\title[Cross-correlation of gravitational lensing from DES Y1 data with SPT \& Planck lensing]{Dark Energy Survey Year 1 Results: Cross-correlation between DES Y1 galaxy weak lensing and SPT+\emph{Planck} CMB weak lensing}

\input{authorlist.tex}

\date{\today}

\label{firstpage}


\begin{abstract}
We cross-correlate galaxy weak lensing measurements from the Dark Energy Survey (DES) year-one (Y1) data with a cosmic microwave background (CMB) weak lensing map derived from South Pole Telescope (SPT) and \emph{Planck} data, with an {effective} overlapping area of 1289 deg$^{2}$. With the combined measurements from four source galaxy redshift bins, we reject the hypothesis of no lensing with a significance of {$10.8\sigma$}. When employing angular scale cuts, this significance is reduced to $6.8\sigma$, which remains the highest signal-to-noise measurement of its kind to date. We fit the amplitude of the correlation functions while fixing the cosmological parameters to a fiducial $\Lambda$CDM model, finding $A = 0.99 \pm 0.17$.  We additionally use the correlation function measurements to constrain shear calibration bias, obtaining constraints that are consistent with
previous DES analyses. Finally, when performing a cosmological analysis under the $\Lambda$CDM model, we obtain the marginalized constraints of {$\Omega_{\rm m}=0.261^{+0.070}_{-0.051}$} and {$S_{8}\equiv \sigma_{8}\sqrt{\Omega_{\rm m}/0.3} = 0.660^{+0.085}_{-0.100}$}. These measurements are used in a companion work that presents cosmological constraints from the joint analysis of two-point functions among galaxies, galaxy shears, and CMB lensing using DES, SPT and \emph{Planck} data. 
\end{abstract}

\preprint{DES-2018-0349}
\preprint{FERMILAB-PUB-18-513-AE}

\maketitle




\section{Introduction}
\label{sec:intro}

As a photon from a distant source travels through the Universe, its path is perturbed by the gravitational potential of large-scale structure, an effect known as gravitational lensing (for a review see e.g. \cite{Bartelmann2001}). The observed amplitude of the perturbations to the photon's trajectory depends on both the matter distribution and geometry of the Universe, making gravitational lensing a powerful cosmological probe.  Furthermore, because these perturbations are induced by gravitational effects, they are sensitive to all forms of matter, including  dark matter, which is difficult to probe by other means. The use of gravitational lensing to constrain cosmology has developed rapidly over the last decade \citep{Fu2008,Jee2012,Heymans2012,Abbott2015,Kilbinger15,Jee2016,Hildebrandt2017,Planck:lensing,Simard17} due to improvements in instrumentation and modeling, and increases in the cosmological volumes probed by surveys \citep{troxel17,des17cosmo}.

In this study, we use two sources of photons to measure the effect of gravitational lensing: distant galaxies and the cosmic microwave background (CMB).  Gravitational lensing caused by the large-scale distribution of matter distorts the apparent shapes of distant galaxies; similarly, gravitational lensing distorts the observed pattern of temperature fluctuations on the CMB last scattering surface. These  distortions are expected to be correlated over the same patch of sky since the CMB photons pass through some of the same intervening gravitational potentials as the photons from distant galaxies.  The two-point correlation between the galaxy lensing and CMB lensing fields can therefore be used as a cosmological probe.

Several features of the cross-correlation between galaxy lensing and CMB lensing make it an appealing cosmological observable.  First, unlike two-point correlations between galaxies and lensing, the lensing-lensing correlation considered here has the advantage that it is not sensitive to difficult-to-model effects such as galaxy bias \citep{Benson:2000}.  Second, since it is a cross-correlation between two independently measured lensing fields from datasets of completely different nature, it is expected to be relatively robust to observational systematics.  For instance, systematics associated with galaxy shape measurement, like errors in the estimate of the point spread function, will have no impact on the inference of CMB lensing.  Third, the use of the CMB lensing field provides sensitivity to the distance to the last scattering surface; the large distance to the last scattering surface in turn provides a long lever arm for constraining cosmology.

Measurement of the two-point correlation between galaxy lensing and CMB lensing was first reported by \cite{hand15} using CMB lensing measurements from the Atacama Cosmology Telescope \citep{fowler07} and galaxy lensing measurements from the Canada-France-Hawaii Telescope Stripe-82 Survey \citep{erben13}.  Several subsequent measurements were made by \cite{liu15} ($\planck$ CMB lensing + CFHTLens galaxy lensing), \cite{kirk16} ($\planck$ and SPT CMB lensing + DES-SV galaxy lensing), \cite{hd16} ($\planck$ CMB lensing + CHTLenS and RCSLenS galaxy lensing), and \cite{hd17} ($\planck$ CMB lensing + KiDS-450 galaxy lensing).  

Here we measure the correlation between CMB lensing and galaxy lensing using CMB data from the South Pole Telescope (SPT) and \planck, and galaxy lensing data from year-one (Y1) observations of the Dark Energy Survey (DES; \cite{des2005}).  We perform a number of robustness checks on the measurements and covariance estimates to show that there is no evidence for significant systematic biases in the measurements over the range of angular scales that we include in the model fits.

The measurements presented here represent the highest signal-to-noise constraints on the cross-correlation between galaxy lensing and CMB lensing to date.  We use the measured correlation functions to place constraints on cosmological parameters (in particular $\omm$ and $S_{8}=\sigma_{8}\sqrt[]{\mathstrut \Omega_{\rm m}/0.3}$).  The cosmological constraints obtained here are complementary to those from DES-Y1 galaxy clustering and weak lensing \citep{des17cosmo}, which are sensitive to somewhat lower redshifts. 

This work is part of a series of four papers that use cross-correlations between DES data and CMB lensing measurements to constrain cosmology:
\begin{itemize}
\item[-] Measurement of correlation between galaxy lensing and CMB lensing (this paper)
\item[-] Measurement of correlation between galaxies and CMB lensing \citep{NKpaper}
\item[-] Methodology for analyzing joint measurements of correlations between DES data and CMB lensing \citep{5x2methods}
\item[-] Results of joint analysis of correlations between DES data and CMB lensing \citep{5x2key}.
\end{itemize}
The main goal of this work is to present the measurement of the correlation between galaxy lensing and CMB lensing, and to subject this measurement to robustness tests.  Consequently, we keep discussion of the cosmological modeling brief and refer the readers to \cite{5x2methods} for a more in depth discussion of the cosmological modeling used in these papers.

This work is organized as follows. In Sec. \ref{sec:theory} we present the theoretical background of the analysis and the required formalism used throughout the analysis. We describe the data products used in Sec. \ref{sec:data} and the methodology used to make the measurements in Sec. \ref{sec:method}. The results are presented in Sec. \ref{sec:results_measurements}, while the cosmological parameter fits are shown in Sec. \ref{sec:results_fits}. { Finally, we present our conclusions in Sec. \ref{sec:conclusion}}.

\section{Theory}
\label{sec:theory}

We are interested in the cross-correlation between CMB lensing and galaxy lensing.  CMB lensing is typically measured in terms of the spin-0 lensing convergence, $\kappa$, which is proportional to a (weighted) integral along the line of sight of the matter density \citep{Lewis2006}.  Galaxy lensing, on the other hand, is most easily measured via the spin-2 \text{shear} field, $\gamma$, by measuring shapes of many galaxies.  The $\gamma$ and $\kappa$ signals are related, and one could in principle convert from $\gamma$ to $\kappa$ \citep[e.g.][]{KS93}.  However, the conversion process is lossy, and not necessary for our purposes since we can directly correlate $\kappa$ and $\gamma$. The galaxy shear signal is estimated from the coherent distortion of the shapes of galaxies.  In this analysis, we measure the correlation of the CMB lensing convergence, $\kcmb$, with the {\it tangential} component of the galaxy shear, $\gamma_{\rm t}$ (i.e. the component orthogonal to the line connecting the two points being correlated).  The advantages of using $\gamma_{\rm t}$ are that it is trivially computed from the observed shear and that $\gamma_{\rm t}$ is expected to be  robust to additive systematics in the shear measurement process.\footnote{Any additive systematic affecting the shear measurements that is constant over scales of interest will vanish when averaging $\gamma_{\rm t}$ over azimuthal angle; this is not true for the Cartesian components of $\gamma$.}  This approach was recently used by \cite{hd16}, who found it to yield higher signal-to-noise than alternative approaches; the same approach was also taken by \cite{Singh:2017}. 

To quantify the correlation between CMB lensing and galaxy lensing, we use the angular two-point function, $w^{\gammat\kcmb}(\theta)$.  To model this correlation, we begin by calculating the theoretical cross-power spectrum between the CMB lensing convergence and the galaxy lensing convergence, $\kcmb$ and $\kgal$, which we denote with $C^{\kgal\kcmb}(\ell) $.  In harmonic-space and using the Limber approximation \citep{limber53, Loverde2008}, we have:
\begin{align}\label{eq:xspec}
C^{\kgal^{i}\kcmb}(\ell) &=\int_{0}^{\chi_{*}}\frac{d\chi}{\chi^2} q_{\kgal^{i}}(\chi)q_{\kcmb}(\chi)P_{\rm NL}\left(k=\frac{\ell+\frac{1}{2}}{\chi},z(\chi)\right),\\
q_{\kappa_{\rm s}^i} &= \frac{3 \Omega_{\rm m} H_0^2}{2c^2}\frac{\chi}{a(\chi)}\int_{\chi}^{\chi_{\rm h}} d\chi' \frac{n^i_{\rm s}(z(\chi')) \frac{dz}{d\chi'}}{\bar{n}_{\rm s}^i} \frac{\chi' - \chi}{\chi'},\\
q_{\kappa_{\rm CMB}} (\chi) &= \frac{3\Omega_{\rm m}H_0^2 }{2c^2}\frac{\chi}{a(\chi)}  \frac{\chi_* - \chi}{\chi_*}. \label{eq:weight_cmbkappa}
\end{align}
Here, $\chi$ is the comoving distance, $\chi_{*}$ is the comoving distance to the last scattering surface, $a(\chi)$ is the cosmological scale factor at distance $\chi$, $n_{\rm s}^{i}(z)$ is the redshift distribution of the source galaxies in the $i$th redshift bin, $\bar{n}_{\rm s}^i=\int dz\ n_{\rm s}^{i}(z)$ is the angular number density in this redshift bin, and $P_{\rm NL}(k,z)$ is the non-linear matter power spectrum at wavenumber $k$ and redshift $z$.  We calculate $P_{\rm NL}$ using the Boltzmann code CAMB\footnote{See \texttt{camb.info}.} \citep{Lewis00,Howlett:2012} with the Halofit extension to nonlinear scales \citep{Smith03,Takahashi:2012} and the \cite{Bird:2002} neutrino extension. 

The harmonic-space cross-spectrum between the CMB and galaxy convergences can be transformed to a position-space correlation function by taking the Hankel transform
\begin{equation}
\label{eq:corr_model}
w^{\gammat^{i}\kappa_{\rm CMB }}(\theta)=\int_{0}^{\infty}\frac{\ell d\ell}{2\pi} C^{\kgal^{i}\kcmb}({\ell}) J_{2}(\ell\theta)F(\ell),
\end{equation}
where $J_{2}$ is the second order Bessel function of the first kind and $F(\ell)$ describes filtering that is applied to the CMB lensing map (see \S\ref{sec:data}).  We set 
\begin{equation}
F(\ell) = \left\{\begin{array}{lr}
        \exp (-\ell(\ell + 1)/\ell_{\rm beam}^2), & \text{for } 30 < \ell < 3000\\
        0, & \text{otherwise,}
        \end{array}\right.
\end{equation}
with $\ell_{\rm beam} \equiv \sqrt{16 \ln 2}/\theta_{\rm FWHM} \approx 2120$, where $\theta_{\rm FWHM} = 5.4'$.  The low and high-pass $\ell$ cuts are imposed to reduce biases in the CMB lensing map; the Gaussian smoothing is applied to ensure that large oscillations are not introduced when transforming from harmonic to position-space {(i.e. aliasing from band limited measurements)}. 

\section{Data}
\label{sec:data}

\subsection{Galaxy weak lensing}
DES is an optical galaxy survey conducted using the 570 Megapixel DECam instrument \citep{Flaugher2015} mounted on the Blanco Telecope at the Cerro Tololo Inter-American Observatory (CTIO) located in Chile. In this analysis, we use the Y1 data that are based on observation runs between August 2013 and February 2014 \citep{Drlica-Wagner17}. We only use the data in the area overlapping with the SPT footprint\footnote{DES-Y1 data also covers the SDSS Stripe-82 region, though the cosmology analysis focuses on the SPT region.}; the overlap area is approximately 1289 $\sqdeg$ between $-60^{\circ}< {\rm Dec.}<-40^{\circ}$, after applying a mask to remove poorly characterized regions.

Two independent shape measurement algorithms --- \metacal\ and \imshape\,  --- were used to generate two different shear catalogs from DES-Y1 data. These algorithms and the corresponding catalogs are described in detail in \cite{Zuntz17}. In this analysis, we only consider the \metacal\ shear estimates because of the higher signal-to-noise ratio of that catalog.

$\metacal$ \citep{huff17,sheldon17} is a recently developed technique for measuring galaxy shears that uses the data itself for calibration, rather than relying on external image simulations. The methodology has been demonstrated to yield a multiplicative shear bias below $10^{-3}$ on simulations with galaxies of realistic complexity \citep{sheldon17}. Briefly, \metacal\ performs shear calibration by applying artificial shears to the observed galaxy images and measuring the response of the shear estimator.  The shear catalog used in this work was based on jointly fitting images in three bands ($riz$). 

The full $\metacal$ catalogue is split into 4 photometric redshift bins: $0.20<z<0.43$, $0.43<z<0.63$, $0.63<z<0.90$, $0.90<z<1.30$, where $z$ is the mean of the estimated redshift probability distribution for each galaxy and the binning is chosen to be consistent with that used in \cite{des17cosmo}. The redshift distributions, $n_{\rm s}^{i}(z)$, for each of the samples were estimated using the \textsc{BPZ} code \citep{Benitez2000}. Detailed validation of these distributions can be found in \cite{Hoyle2017,Gatti2017,Davis2017}.  We also checked that using an independent $n_{\rm s}^{i}(z)$ estimation from the high quality COSMOS2015 photometric redshift catalog \citep{Laigle2016,Hoyle2017} results in negligible change in the final cosmological constraints.

To avoid implicit experimenter bias, the measurements were blinded while most of the analysis was being performed. The measurements were not compared with theoretical predictions and the axes were removed prior to unblinding. For cosmological parameters estimations, the contours were shifted, and the axes were removed.

\begin{figure}
\begin{center}
\includegraphics[width=1.00\linewidth]{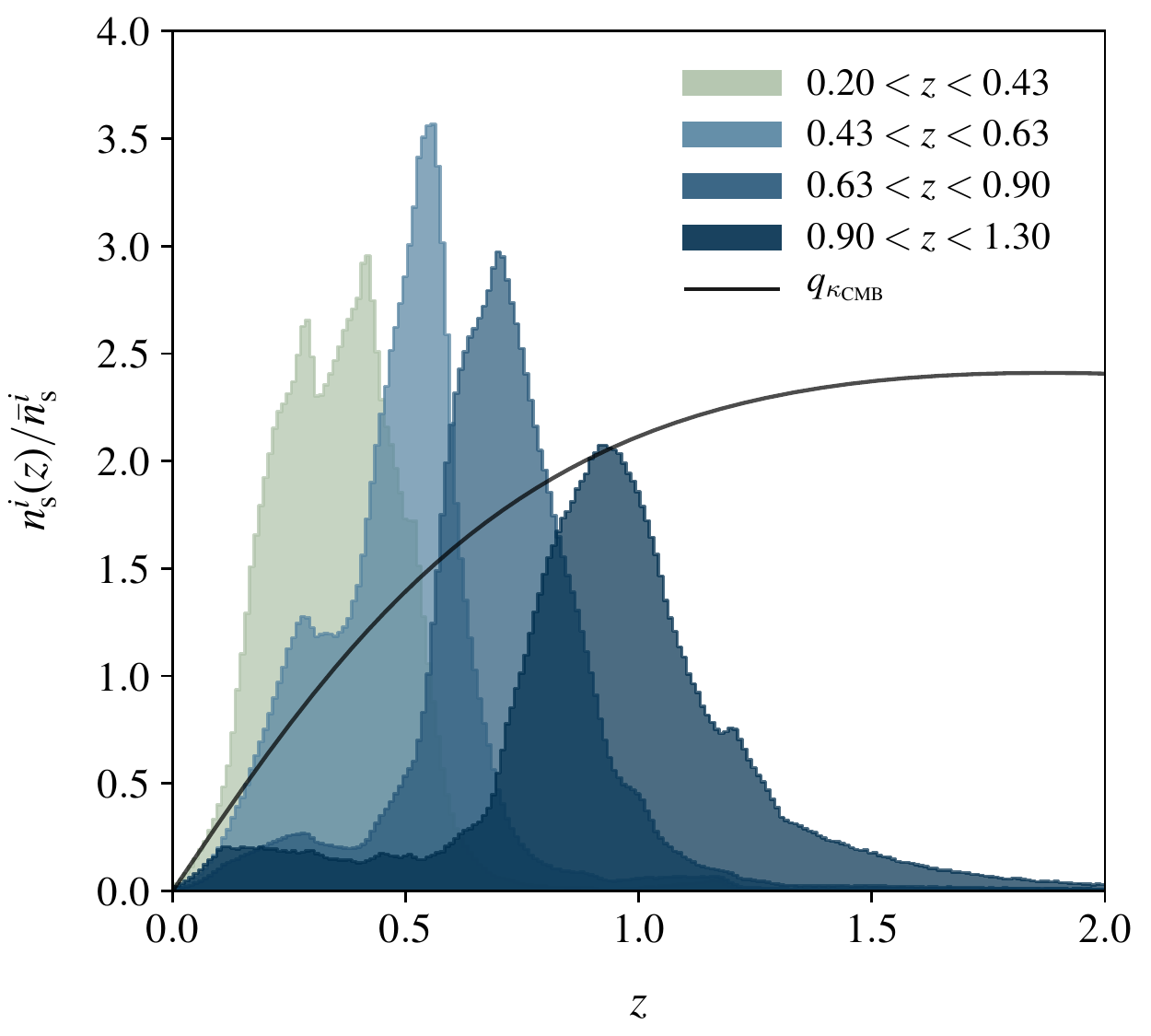}
\caption{Redshift distribution of galaxies $n_{\rm s}^{i}(z)$ for the 4 tomographic bins for $\metacal$. The black line shows the CMB lensing kernel.}
\label{fig:footprint}
\end{center}
\end{figure}

\subsection{CMB lensing map}

We use the CMB weak lensing map described in \cite{omori17}, which was created from a combination of the SPT and \emph{Planck} CMB temperature data.  Details of the $\kappa_{\rm CMB}$ procedures used to create the map can be found in \cite{omori17}; we provide a brief overview below.

The lensing map is derived from a minimum-variance combination of SPT 150\ GHz and \emph{Planck} 143\ GHz temperature maps over the SPT-SZ survey region ($20^{\rm h}$ to $7^{\rm h}$ in right ascension 
and from $-65^\circ$ to $-40^\circ$ in declination).  By combining SPT and \Planck maps in this way, the resultant temperature map is sensitive to a greater range of modes on the sky than either experiment alone.  Modes in the temperature maps with $\ell>3000$ are removed to avoid systematic biases due to astrophysical foregrounds such as the thermal Sunyaev-Zel'dovich effect (tSZ) and the cosmic infrared background (CIB) \citep{vanengelen15}, whereas modes with $\ell < 100$ are removed to reduce the effects from low-frequency noise. The quadratic estimator technique \citep{okamoto03} is used to construct a (filtered) estimate of $\kappa_{\rm CMB}$.  Simulations are used to remove the mean-field bias and {to calculate the response function which is used} to properly normalize the amplitude of the filtered lensing map.

The output lensing convergence map is filtered {further} to remove modes with $\ell < 30$ and $\ell > 3000$ and is smoothed with a Gaussian beam with full width at half maximum of $5.4'$.  Point sources {(dusty-star forming and radio galaxies)} with flux density above 6.4 mJy in the 150 GHz band are masked with apertures of  $r=3',6',9'$ depending the brightness of the point source.  Additionally, in order to reduce contamination of the $\kappa_{\rm CMB}$ map by thermal Sunyaev-Zel'dovich (tSZ) signal, we apply a mask to remove clusters detected at signal-to-noise ${\rm S/N}>5$ in the SPT CMB maps, and DES $\redmapper$ clusters with richness $\lambda>80$; these clusters are masked with an aperture of $r=5'$.  The effectiveness of this masking at reducing tSZ contamination was investigated in \cite{5x2methods}. Such masking could in principle induce a bias because clusters are associated with regions of high lensing convergence.  However, it was shown in \cite{5x2methods} that this bias is negligible {due to the small fraction of area masked relative to the total area used}.\footnote{Less than 1\% of the survey area is lost by applying a mask that removes 437 clusters.} 

{The effect of the uncertainty on the calibration of the CMB temperature was investigated in \cite{omori17}, and it was found to be at most $0.20\sigma$ of the statistical uncertainty when the calibration is conservatively varied by 1\% (although it is known to better than 1\% as noted in \cite{hou18}).}

\section{Methods} 
\label{sec:method}

\subsection{Two-point measurement}
Our estimator for the angular correlation function at the angular bin specified by angle $\theta_{\alpha}$ is
\begin{align}
w^{\gamma_{\rm t}\kappa_{\rm CMB}}(\theta_{\alpha})=&\frac{\sum_{i=1}^{N_{\rm pix}} \sum_{j=1}^{N_{\rm gal}} f^{i}_{\kappa}  \kappa^{i}_{\rm CMB}e_{\rm t}^{ij}\Theta_{\alpha}(\hat{\theta}^i - \hat{\theta}^j)}{s (\theta_{\alpha}) \sum f^{i}_{\kappa}},
\label{eq:measurement}
\end{align}
where the sum in $i$ is over all pixels in the CMB convergence map, the sum in $j$ is over all source galaxies, and $\hat{\theta}$ represents the direction of the $\kcmb$ pixels or source galaxies.  $e_{\rm t}^{ij}$ is the component of the corrected ellipticity oriented orthogonally to the line connecting pixel $i$ and source galaxy $j$ \citep[see e.g.][]{prat17}.  The $\kcmb$ value in the pixel is $\kcmb^i$ and  $f^{i}_{\kappa}$ is the associated pixel masking weight, which takes a value between zero and one (i.e. zero if the pixel is completely masked).  The function $\Theta_{\alpha}(\theta)$ is an indicator function that is equal to unity when the angular separation between $\hat{\theta}_i$ and $\hat{\theta}_j$ is in the angular bin specified by $\theta_{\alpha}$, and zero otherwise.  Finally, $s (\theta_{\alpha})$ is the $\metacal$ response, which can be estimated from the data using the procedure described in \cite{Zuntz17}.  We find that $s (\theta)$ is approximately constant over the angular scales of our interest, but different for each redshift bin.  We evaluate the estimator in Eq.~\ref{eq:measurement} using the $\treecorr$ package.\footnote{\url{https://github.com/rmjarvis/TreeCorr}}

We perform the $w^{\gammat\kcmb}(\theta)$ measurements in 10 logarithmic bins over the angular range $2.5'<\theta<250'$.  Later we remove a sub-range of these scales in the likelihood analysis, where the scale cuts are determined such that they prevent known sources of systematic error from biasing cosmological constraints (see Sec. \ref{sec:scalecuts}).

\subsection{Modelling of systematic effects in galaxy shear measurements}\label{sec:modelling}

Eq.~\ref{eq:corr_model} forms the basis for our model of the measured correlation functions. We improve on this basic model by also incorporating prescriptions for systematic errors in the estimated shears and redshift distributions of the galaxies.  We describe these models briefly below. For more details, readers should refer to \cite{5x2methods} and \cite{krause17}.  The computation of the model vectors and sampling of parameter space is performed using $\textsc{CosmoSIS}$ \citep{Zuntz:2015,Lewis00,Smith03,Bridle07,Kilbinger09,Howlett12,Kirk12,Takahashi12}.

\subsubsection{Photometric redshift bias}

The inference of the redshift distribution, $n_{\rm s}^{i}(z)$, for the source galaxy sample is potentially subject to systematic errors.
Following \cite{5x2methods}, \cite{krause17} and related past work \citep{Abbott2015,Joudaki2017,Bonnett2016,Becker2016}, we account for these potential systematic errors in the modeling by introducing a photometric redshift bias parameter which shifts the assumed $n_{\rm s}^{i}(z)$ for the source galaxies.  That is, the true redshift distribution for the $i$th source galaxy bin, $n_{{\rm s}, \mathrm{unbiased}}^{i}(z)$, is related to the observed redshift distribution, $n_{\rm s}^{i}(z)$, via:
\begin{equation}
n_{{\rm s},\mathrm{unbiased}}^{i}(z)=n_{{\rm s}}^{i}(z-\Delta_{z,{\rm s}}^{i}),
\end{equation} 
where $\Delta_{z,{\rm s}}^{i}$ is the redshift bias parameter, which is varied independently for each source galaxy redshift bin. 

Priors on the $\Delta_{z,{\rm s}}^{i}$ are listed in Table~\ref{table:prior}. The $\Delta_{z,{\rm s}}^{i}$ values for the three lowest redshift bins were obtained by cross-correlating the source galaxy sample with $\redmagic$ Luminous Red Galaxies (LRGs) \citep{Rozo2016redmagic}, which have well characterized redshifts. The $\Delta_{z,{\rm s}}^{i}$ value for the highest redshift bin comes from comparing $n_{\rm s}^{i}(z)$ derived from BPZ and the COSMOS2015 catalog. The derivation of these priors is described in \cite{Hoyle2017}, with two other supporting analyses described in \cite{Gatti2017} and \cite{Davis2017}.

\subsubsection{Shear calibration bias}

In weak lensing, one estimates galaxy shapes, or ellipticities using a suitably chosen estimator. These estimators are often biased and need to be \textit{calibrated} using either external image simulations (e.g. the $\imshape$ method) or manipulation of the data itself (e.g. the $\metacal$ method). The shear calibration bias refers to the residual bias in the shear estimate after the calibration process, or the uncertainty in the calibration process. In particular, we are mainly concerned about the multiplicative bias in the shear estimate, which can arise from failures in the shape measurements, stellar contamination in the galaxy sample, false object detection and selection bias \citep{heymans06,HTBJ2006}.

Following \cite{5x2methods} and \cite{krause17}, we parameterize this systematic error in shear calibration with  a single multiplicative factor, ($1+m_{i}$), for each redshift bin $i$. With this factor, the observed correlation function becomes:
\begin{equation}
\hspace{0.5cm}w_{{\rm obs}}^{\gammat\kcmb}(\theta)=(1+m^{i})w_{{\rm true}}^{\gammat\kcmb}(\theta)\hspace{1cm} i\in\{1,2,3,4\}.
\end{equation}
We let the bias parameter for each redshift bin vary with a Gaussian prior listed in Table~\ref{table:prior} based on \cite{Zuntz17}. 

\subsubsection{Intrinsic alignment}\label{sec:modelling_IA}

In addition to the apparent alignment of the shapes of galaxies as a result of gravitational lensing, galaxy shapes can also be {\it intrinsically} aligned as a result of their interactions with the tidal field from nearby large scale structure.  The intrinsic alignment (IA) effect will impact the observed correlation functions between galaxy shear and $\kcmb$ \citep{Hall2014,Troxel2014}. The impact of IA can be modeled via:
\begin{equation}
C_{{\rm obs}}^{\kgal\kcmb}(\ell)=C_{{\rm true}}^{\kgal\kcmb}(\ell)-C^{\kcmb {\rm I}}(\ell),
\end{equation}
where $C^{\kcmb {\rm I}}(\ell)$ is calculated in a similar way as Eq. \ref{eq:xspec}, but with replacing the galaxy lensing kernel with: 
\begin{equation}
W^{\rm I}(\chi)=A(\chi(z))\frac{C_{1}\rho_{\rm crit}\Omega_{\rm m}}{\mathcal{D}(z)}\frac{n_{\rm s}^i(z(\chi))}{\bar{n}_{\rm s}^i} \frac{dz}{d\chi},
\end{equation}
where $\mathcal{D}(z)$ is the linear growth function. Here we have employed the non-linear linear alignment model (see \cite{Bridle07} for details) and included the redshift evolution of the IA amplitude via
\begin{equation}
A(\chi(z))=A^{\rm IA}\left(\frac{1+z}{1+z_{0}}\right)^{\eta^{\rm IA}},
\end{equation}
We use fixed values $z_{0}=0.62$, $C_{1}\rho_{\rm crit}=0.0134$, while letting $A^{\rm IA}$ and $\eta^{\rm IA}$ vary, as done in \cite{des17cosmo}. 

\begin{table}
\footnotesize
\centering

\footnotesize
\centering

 \begin{tabular}{ccc}
\hline
\hline
\textsc{parameter} & \textsc{fiducial} &\textsc{prior}\\\hline
\textsc{cosmology}\\
$\Omega_{\rm m}$ & 0.309 &  [0.1, 0.9] \\ 
$A_\mathrm{s}/10^{-9}$ &$2.14$ &  [$0.5$, $5.0$]  \\ 
$n_{\rm s}$ & 0.967 & [0.87, 1.07]  \\
$\w$ &  -1.0 &   \textsc{fixed}   \\
$\omb$ &  0.0486 &  [0.03, 0.07]  \\
$h_0$  & 0.677 &  [0.55, 0.91]   \\
$\Omega_\nu h^2$ & $6.45\times 10^{-4}$ & [0.0006,0.01] \\
$\Omega_\mathrm{K}$ & $0$ & \textsc{fixed} \\
$\tau$ & $0.066$ & \textsc{fixed}\\

\textsc{shear calibration bias} & \\

$m^{1}$ & 0.010 & $(0.012,0.023)$ \\
$m^{2}$ & 0.014 & $(0.012,0.023)$ \\
$m^{3}$ & 0.006 & $(0.012,0.023)$ \\
$m^{4}$ & 0.013 & $(0.012,0.023)$ \\[0.15cm]

 \textsc{intrinsic alignment}  & \\
$A^{\rm IA}$& 0.44  & $[-5,5]$\\
$\eta^{\rm IA}$& -0.67 &$[-5,5]$\\
$z_{0}$ & $0.62$ & \textsc{fixed}\\[0.15cm]

 \textsc{source photo-$z$ error} & \\
$\Delta^{1}_{z,{\rm s}}$ & -0.004 &(-0.001,0.016)\\
$\Delta^{2}_{z,{\rm s}}$ & -0.029 &
(-0.019,0.013)\\
$\Delta^{3}_{z,{\rm s}}$ & 0.006 &
(0.009,0.011) \\
$\Delta^{4}_{z,{\rm s}}$ & -0.024 &
(-0.018,0.022) \\[0.15cm]\hline
\end{tabular}
\caption{
The fiducial parameter values\footnote{We use the \textit{Planck} \textsc{TT,TE,EE}+\textsc{lensing+ext} best-fit values from \citep{planck2015xiii} for the cosmological parameters and the marginalized 1D peaks for the DES nuisance parameters from the DES-Y1 joint analysis \citep{des17cosmo}.}\ and priors for cosmological and nuisance parameters used in this analysis. Square brackets denote a flat prior over the indicated range, while parentheses denote a Gaussian prior of the form $\mathcal{N}(\mu,\sigma)$.}
\label{table:prior}
\end{table}

\subsection{Covariance}
The covariance matrix of $w^{\gammat\kcmb}(\theta)$ is computed analytically, using  the halo-model to estimate the non-Gaussian contributions.  Details of the covariance calculation can also be found in \cite{5x2methods} and \cite{krause17}. {However, we make a small modification in calculating the noise-noise covariance term, which we measure by cross-correlating $\kcmb$ noise and rotated galaxy shears. This modification is needed to incorporate the geometry of the mask, which the analytic covariance neglects, and this correction increases the covariance by  $\sim30\%$.} We compare the theoretical estimate of the covariance to an estimate of the covariance derived from the data in Sec.  \ref{sec:covariance_test}. 

\subsection{Angular scale cuts}\label{sec:scalecuts}

There are several effects that may impact the observed correlation functions that we do not attempt to model.  As shown in \cite{5x2methods}, the most significant unmodelled effects for the analysis of $w^{\kcmb\gamma_{\rm t}}$ are bias in $\kappa_{\rm CMB}$ due to the thermal Sunyaev-Zel'dovich (tSZ) effect, and the impact of baryonic effects on the matter power spectrum.  To prevent these effects from introducing systematic errors into our cosmological constraints, we exclude the angular scales from our analysis that are most impacted.  The tSZ bias is greatest at approximately $3-5$~Mpc and slowly decreases at both smaller and larger scales. In contrast, the impact of the baryonic effects is greatest below roughly 4~Mpc, and negligible at larger scales (see Fig. 4 of \cite{5x2methods}). Based on these results, it was demonstrated that the impact of the sum of these effects can be mitigated by excluding small scales from the analysis.

In this study we adopt the scale cuts directly from \cite{5x2methods}. The scale cuts exclude angular bins below 40 arcminutes for the two lowest redshift bins, and scales below 60 arcminutes for the two highest redshift bins.\footnote{These angular scales cuts are applied to the two-point correlation measurement between galaxy weak lensing and the CMB lensing map, not the temperature map that is used to reconstruct the lensing map.} {For the angular scales beyond the scale cuts, the baryonic effects have negligible impact to our measurements, while the tSZ effect still remains. We quantify this residual bias in Sec.~\ref{sec:sys}. }

We note that the scale cut choices made in this analysis were motivated from consideration of the full \5x2pt data vector, and not from consideration of $w^{\gammat\kcmb}(\theta)$ alone.  This choice was made because one of the main purposes of this work is to provide the measurements of $w^{\gammat\kcmb}(\theta)$ that will be incorporated into the companion analysis of \cite{5x2key}.  Since the other four two-point functions also contribute some potential bias in the \5x2pt analysis, the scale cut choice adopted here is conservative for the analysis of $w^{\gammat\kcmb}(\theta)$ alone.

\section{Measurement}
\label{sec:results_measurements}

\begin{figure}
\begin{center}
\includegraphics[width=1.0\linewidth]{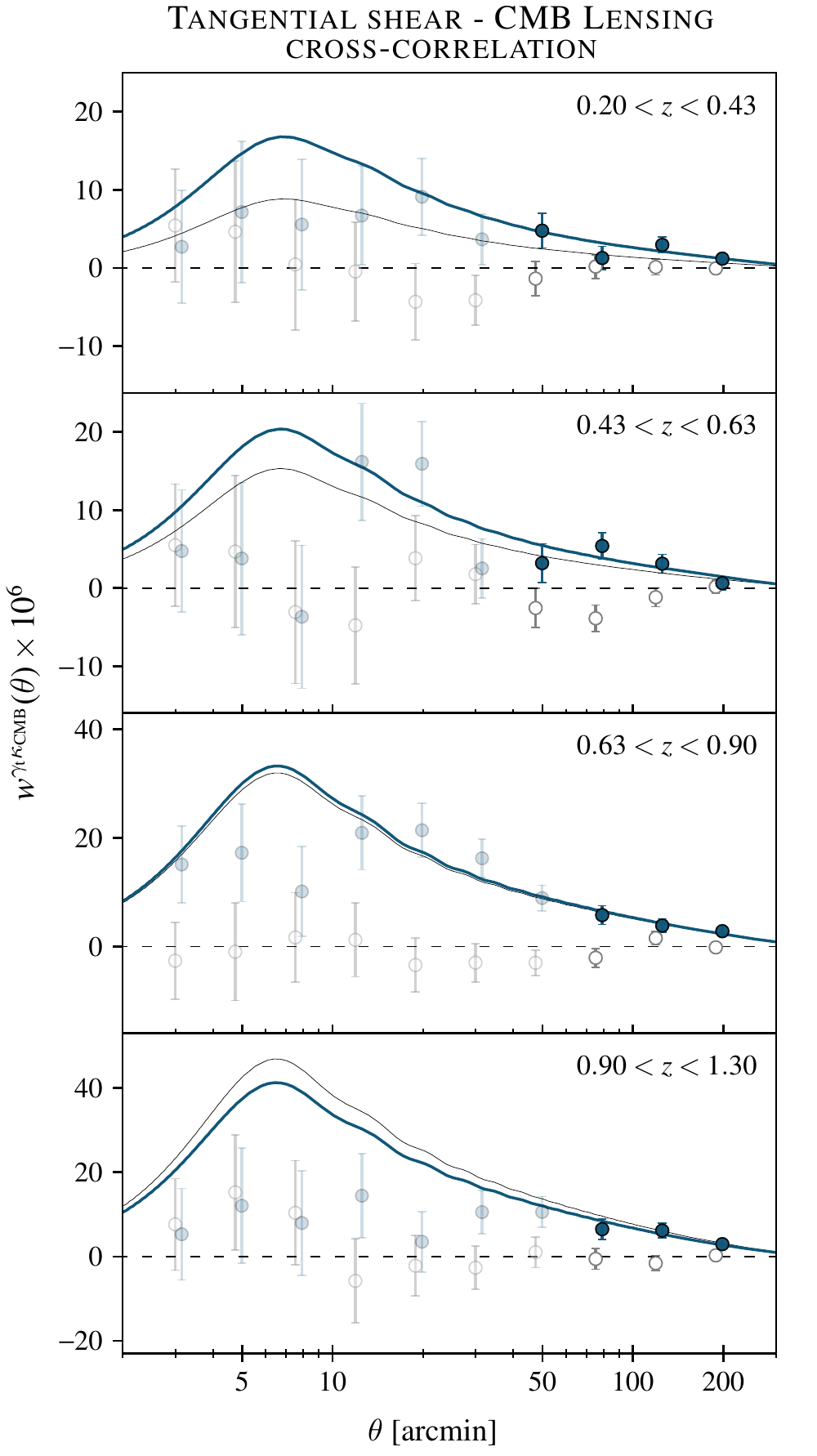}
\caption{Measurements of $w^{\gamma_{\rm t}\kappa_{\rm CMB}}(\theta)$ (filled circles) and $w^{\gammax\kappa_{\rm CMB}}(\theta)$ (open circles)  using \metacal\ shear estimates and the SPT+\emph{Planck} CMB lensing map. The four panels show results for the four source galaxy redshift bin. Faded points are removed from the final analysis due to systematics or uncertainties in the modeling. Also shown  are the theoretical predictions using fiducial cosmology with $A=1$ (black), and with best-fit $A$ (blue), where $A$ in defined in Sec. \ref{sec:amplitude}.}
\label{fig:gammat}
\end{center}
\end{figure}

The measured two-point angular correlation functions, $w^{\gammat\kcmb}(\theta)$, for each of the source galaxy bins are shown in Fig.~\ref{fig:gammat}. For each redshift bin we measure the correlation function in 10 angular bins logarithmically spaced between 2.5 and 250 arcminutes.  We choose this binning to preserve reasonable signal-to-noise in each angular bin, as discussed in \cite{5x2methods}. 

\subsection{Testing the measurements}

\subsubsection{Correlation of $\kappa_{\rm CMB}$ with $\gamma_{\times}$}

When cross-correlating the observed galaxy shears with the $\kappa_{\rm CMB}$ map, we divide the observed shear into a tangential component, $\gammat$, oriented tangentially to the line connecting the two points being correlated, and a cross component, $\gamma_{\times}$, which is parallel to the line connecting the two points. Weak lensing is expected to produce a tangential shear component only, and therefore the presence of a non-zero cross-correlation with the cross-shear component would indicate the presence of systematic errors.

In Fig.~\ref{fig:gammat}, we show the measured cross-correlation between the $\kappa_{\rm CMB}$ maps and the cross-component of the shear ({open points}). As expected, we find that the measured cross-correlation is consistent with zero in all redshift bins. We calculate the $\chi^{2}/\nu$ (where $\nu$ is the number of degrees of freedom) and {probability-to-exceed} (p.t.e.) between the measurement and the null hypothesis (zero cross-correlation) for all redshift bins combined, applying the angular scale cuts described in Sec. \ref{sec:scalecuts}, and find $\chi^{2}/\nu=6.9/14$ and p.t.e = $0.94$, indicating consistency of the cross-shear correlation with zero. {The $\chi^{2}/\nu$ and p.t.e for the individual bins are summarized in Table \ref{table:bestfitA}.}

\subsection{Testing the covariance}
\label{sec:covariance_test}

As mentioned in Sec.~\ref{sec:modelling}, we employ a theoretical covariance matrix (with a small empirical modification) when fitting the measured correlation functions. To test whether the theoretical covariance accurately describes the noise in the measurements, we compare it to an estimate of the covariance obtained using the ``delete-one" jackknife method applied to data. 

To compute the jackknife covariance estimate, we divide the source galaxy samples into $N_{\rm jk} = 100$ approximately equal-area patches. The jackknife estimate of the covariance is then computed as
\begin{equation}
\mathbf{C}^{\rm jackknife}_{ij} = \frac{N_{\rm jk}-1}{N_{\rm jk}} \sum_{k} (d_i^{k} - \bar{d_{i}})(d_j^{k} - \bar{d_{j}}),
\label{eq:jk1}
\end{equation}
where $d^k_i$ is $i$th element of the ${w}^{\gammat\kcmb}(\theta)$ data vector that is measured after excluding the shears in the $i$th patch on the sky, and $\bar{d}$ is 
\begin{equation}
\bar{d}_{i} = \frac{1}{N_{\rm jk}} \sum_k d_{i}^{k}.
\label{eq:jk2}
\end{equation}
We have validated the jackknife approach to estimating the covariance matrix of ${w}^{\gammat\kcmb}(\theta)$ using simulated catalogs. The validation tests are described in Appendix~\ref{sec:jackknife_validation}.

\begin{figure}
\begin{center}
\includegraphics[width=0.48\textwidth]{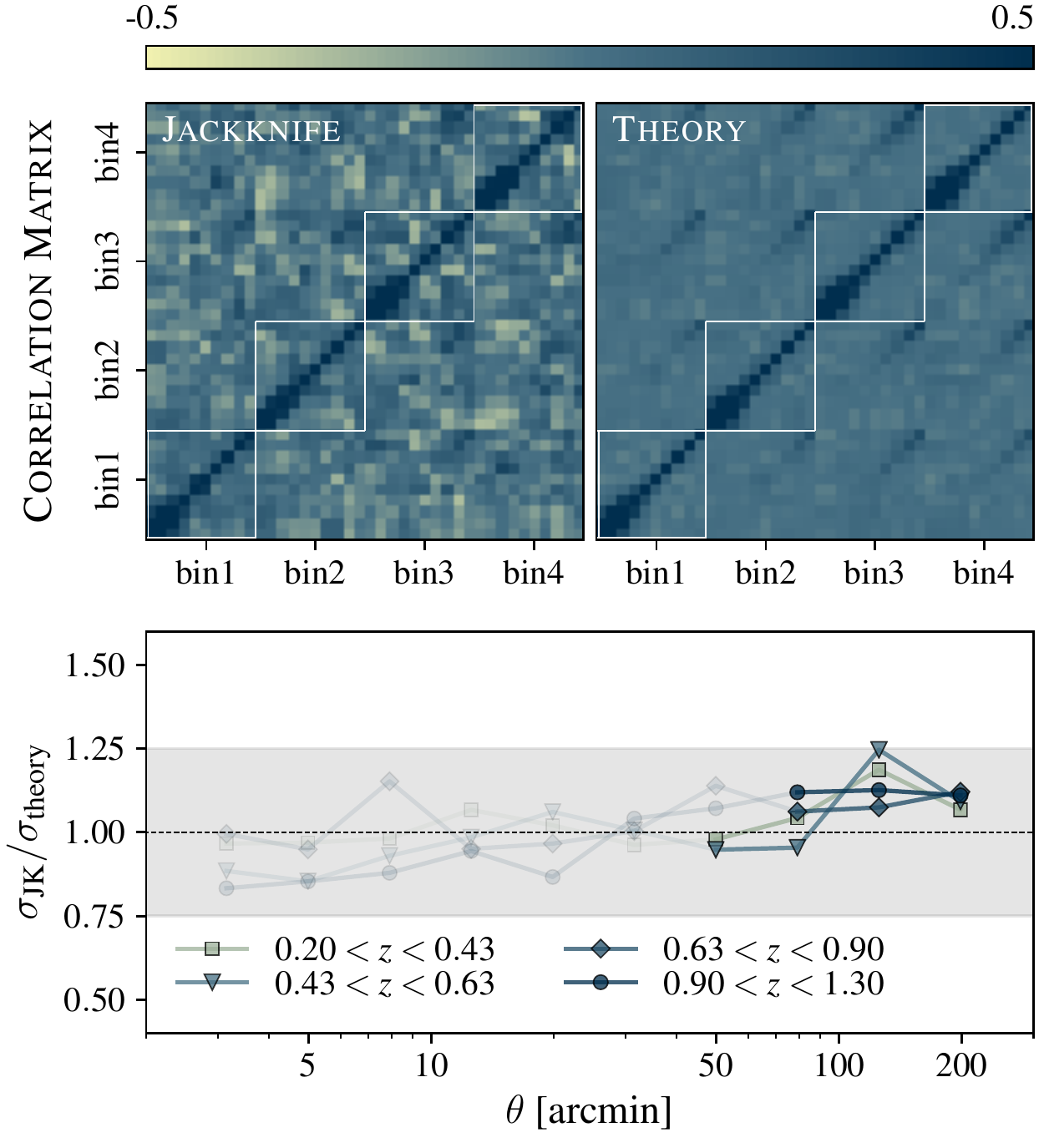}
\caption{The jackknife (upper left) and theory (upper right) correlation matrix ($\mathbf{C}_{ij}/\sqrt{\mathbf{C}_{ii}\mathbf{C}_{jj}}$) for all the redshift bins. Lower panel: ratio of the diagonal component of the covariance matrix for the theory and the jackknife covariance in all redshift bins showing an agreement to within 25\% (shown as the gray band) for all the redshift bins.}
\label{fig:covariance_compare}
\end{center}
\end{figure}



The theoretical and jackknife estimates of the covariance matrix, and the ratio between the diagonal elements of the two  are shown in Fig.~\ref{fig:covariance_compare}. It is clear from the top panels of the figure that the covariance structure of the theoretical covariance agrees qualitatively with the covariance measured from the data. Furthermore, the bottom panel shows that the two covariances agree along the diagonal to better than $25$\%\footnote{25\% is approximately the scatter we see when comparing the covariance computed from many \flask\ (described in Appendix \ref{sec:flask}.) realizations and using the jackknife method on a single \flask\  realization.} across all redshift bins.

\subsection{Estimating the impact of unmodeled systematics}
\label{sec:sys}
While some sources of systematic error are modeled in the analysis (namely photometric redshift and multiplicative shear biases), there are several other potential sources of systematic errors coming from unmodeled effects that could impact the measurement of $w^{\gammat\kcmb}(\theta)$.  Some of these, such as tSZ bias, are minimized with angular scale cuts.  One useful diagnostic to determine the impact of residual systematic biases is to identify the list of external quantities that could directly or indirectly contaminate the signal and cross-correlate them with the measured galaxy shears and CMB convergence. We expect these cross-correlations to be consistent with zero if these external quantities are not introducing significant biases in the measurements.  One example of a quantity that could correlate both with observed  shear and CMB convergence is dust extinction: dust extinction is lower at high galactic latitudes, which is where the density of stars is lowest, and therefore, could result in poor PSF modelling and biased shear estimates in those areas. Meanwhile, dust is one of the foreground component of the CMB temperature measurements, and one can expect potential residuals in a single frequency temperature map. When a contaminated temperature map is passed through the lensing reconstruction pipeline, fluctuations from these foregrounds get picked up as false lensing signal, which will be spatially correlated with the variations in the galaxy shape measurements, and therefore introduce biases in our measurements.
 
We divide potential systematic contaminants into two categories: those that are expected to be correlated with the true (i.e. uncontaminated) $\gamma$ or $\kappa_{\rm CMB}$, and those that are not. For those systematics that are expected to be uncorrelated with the true $\gamma$ and $\kappa_{\rm CMB}$, we estimate the contamination of $w^{\gammat\kcmb}(\theta)$ via 
\begin{equation}
\label{eq:systematic_bias}
w_{\mathcal{S}}(\theta)= \frac{w^{\kappa_{\rm CMB} \mathcal{S}}(\theta) \,  w^{\gamma_{\rm t} \mathcal{S}}(\theta)} {w^{\mathcal{S} \mathcal{S}}(\theta)},
\end{equation}
where $\mathcal{S}$ is the foreground map of interest. This expression captures correlation of the systematic with both $\kappa_{\rm CMB}$ and $\gamma$, and is normalized to have the same units as $w^{\gammat\kcmb}(\theta)$. Unless the systematic map is correlated with both $\gammat$ and $\kappa_{\rm CMB}$, it will not bias $w^{\gammat\kcmb}(\theta)$ and $w_{\mathcal{S}}(\theta)$ will be consistent with zero.

\begin{figure}
\begin{center}
\includegraphics[width=1.00\linewidth]{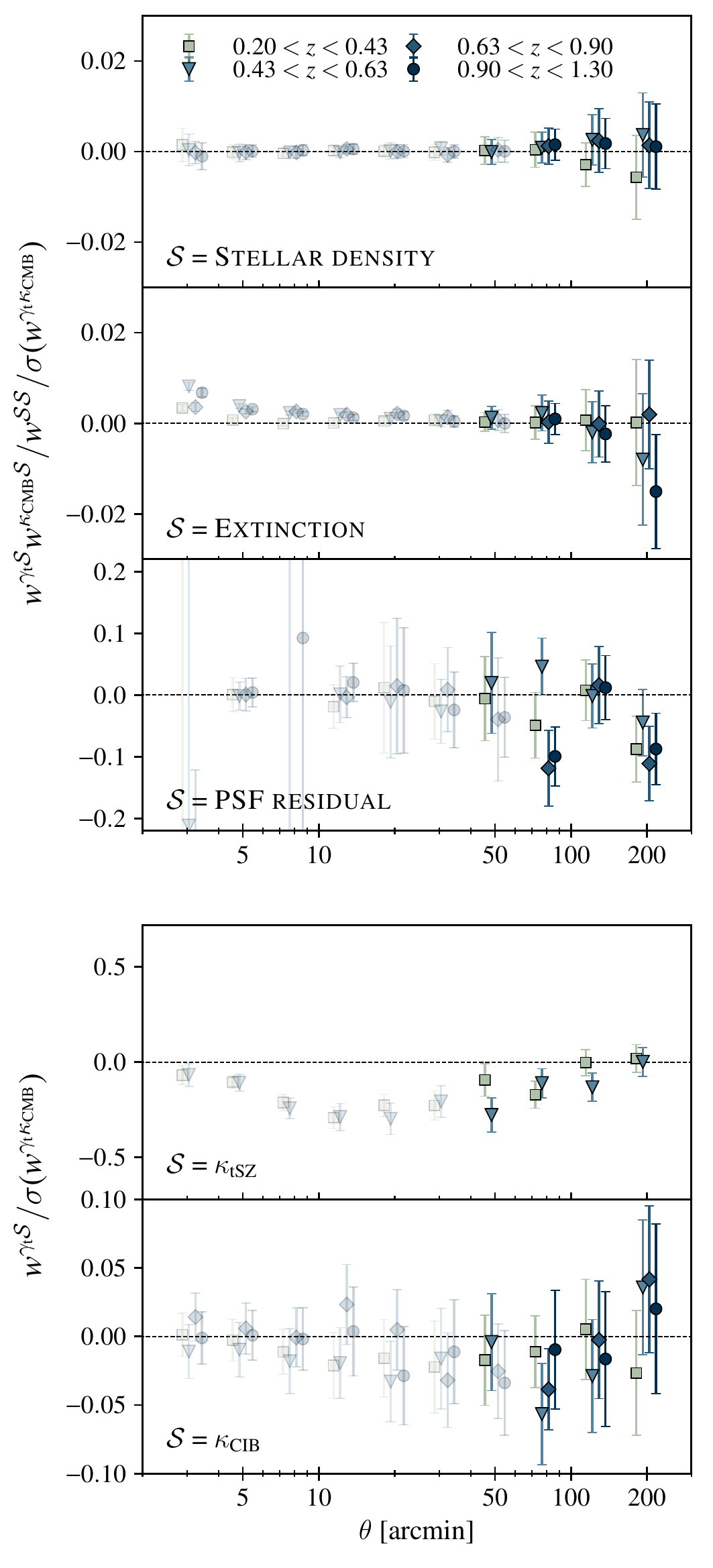}
\caption{Ratios of the estimated systematic biases to $\gammat\kcmb$ from various contaminants to the statistical uncertainties on $\gammat\kcmb$.  We find that all systematics considered result in negligible bias to the $\gammat\kcmb$ measurements.  For the case of PSF residuals, the auto-correlation $w^{\mathcal{S}\mathcal{S}}(\theta)$ of some bins are close to zero, resulting in large error bars for certain bins. As described in the text, contamination from the tSZ effect and the CIB (bottom two panels) must be treated somewhat differently from the other contaminants, since these two potential sources of bias are known to be correlated with the signal.  While we find significant evidence for non-zero $w^{\gamma_{\rm t}\kappa_{\rm tSZ}}(\theta)$, the size of this correlation is small compared to the errorbars on $w^{\gammat\kcmb}(\theta)$, and does not lead to significant biases in cosmological constraints.}
\label{fig:syscrosses}
\end{center}
\end{figure}

We consider three potential sources of systematic error that are expected to be uncorrelated with the true $\gamma$ and $\kappa_{\rm CMB}$: $\gamma^{\rm PSFres}_{\rm t}$ (the residual PSF ellipticity), $E_{B-V}$ (dust extinction) and $\delta_{\rm star}$ (stellar number density). We use the difference between the PSF ellipticity between the truth (as measured from stars) and the model for the PSF residual. Descriptions of the $E_{B-V}$ and $\delta_{\rm star}$ maps can be found in \cite{ElvinPoole2017}. The measured $w_{\mathcal{S}}(\theta)$ for these quantities are plotted in Fig.~\ref{fig:syscrosses} relative to the uncertainties on $w^{\gammat\kcmb}(\theta)$.  The error bars shown are determined by cross-correlating the systematic maps with simulated $\kcmb,\gammat$ maps generated using the $\flask$ package \citep{Xavier:2016}.  For each of the potential systematics considered, we find that the measured $w_{\mathcal{S}}(\theta)$ is much less than the statistical uncertainties on the $w^{\gammat\kcmb}(\theta)$ correlation, implying that there is very little impact from these systematics. 

Astrophysical systematic effects that we expect to correlate with the true $\gamma$ and $\kappa_{\rm CMB}$ must be treated somewhat differently, since in this case, Eq.~\ref{eq:systematic_bias} will not yield the expected bias in $w^{\gammat\kcmb}(\theta)$.  Two sources of potential systematic error are expected to have this property, namely contamination of the $\kappa_{\rm CMB}$ map by tSZ and the cosmic infrared background (CIB).  Since the tSZ and CIB are both correlated with the matter density, these contaminants will be correlated with the true shear and $\kappa_{\rm CMB}$ signals.  For both contaminants, we construct convergence maps of the contaminating fields across the DES patch, which we refer to as $\kappa_{\rm tSZ}$ and $\kappa_{\rm CIB}$.  The estimates of $\kappa_{\rm tSZ}$ and $\kappa_{\rm CIB}$ are generated as described in  \cite{5x2methods}.

We estimate the bias induced to $w^{\gammat\kcmb}(\theta)$ by tSZ and CIB by measuring $w^{\gammat\kappa_{\rm tSZ}}(\theta)$ and $w^{\gammat\kappa_{\rm CIB}}(\theta)$.  These quantities are plotted in Fig.~\ref{fig:syscrosses}, with error bars determined {by measuring the variance between the systematic maps with 100 simulated sky realizations generated using the $\flask$ simulations (see Appendix \ref{sec:flask} for details)}.  We measure a bias over the angular ranges of interest, with a maximum bias\footnote{{\cite{5x2methods} uses theory data vectors and model fits to the measured biases  to calculate similar quantities, from which the scale-cuts are derived. In contrast, the measurements shown in Fig.~\ref{fig:syscrosses} are calculated using the $\ktsz$ map and the galaxy shape catalogs, and therefore includes scatter. Although it may appear as though the scale cuts are removing less biased angular bins, this is primarily due to the scatter in our measurements.}} of $\sim0.30\sigma$ (where $\sigma$ is the expected standard deviation for $w^{\gammat\kcmb}(\theta)$). As shown in \cite{5x2methods}, this level of bias results in a small shift to inferred parameter constraints.

\section{Parameter Constraints}
\label{sec:results_fits}

We assume a Gaussian likelihood for the data vector of measured correlation functions, $\vec{d}$, given a model, $\vec{m}$, generated using the set of parameters $\vec{p}$:
\begin{multline}
\ln \mathcal{L}(\vec{d}|\vec{m}(\vec{p}))=  
-\frac{1}{2} \sum^N_{ij} \left(d_i - m_i(\vec{p})\right) \mathbf{C}^{-1}_{ij} \left(d_j - m_j(\vec{p}) \right),  
\end{multline}
where the sums run over all of the $N$ elements in the data and model vectors. The posterior on the model parameters can be calculated as: 
\begin{equation}
P(\vec{m}(\vec{p})|\vec{d}) \propto \mathcal{L}(\vec{d} | \vec{m}(\vec{p})) P_{\rm prior} (\vec{p}),
\end{equation}
where $P_{\rm prior}(\vec{p})$ is the prior on the model parameters.

In the following sections, we will use this framework to generate parameter constraints in four scenarios, each keeping different sets of parameters free.

We note that we made minor modifications to the analysis after we unblinded the data. We originally computed the constraints on shear calibration and intrinsic alignment parameters fixing the cosmology to the values obtained from DES-Y1 in Sec.~\ref{sec:shearcal} and \ref{sec:IA}. We later allowed the cosmological parameters to vary but combined with the $\planck$ baseline likelihood. Consequently, we also switched to using models generated assuming $\planck$ best-fit values when fitting the correlation amplitudes in Sec.~\ref{sec:amplitude}, so that the same framework is used  throughout the analysis.

\subsection{Amplitude fits}
\label{sec:amplitude}

We first attempt to constrain the amplitude of the observed correlation functions relative to the expectation for the fiducial cosmological model summarized in Table~\ref{table:prior}.  The fiducial cosmological parameters are chosen to be the best-fitting parameters from the analysis of CMB and external datasets in \cite{planck2015xiii}; and nuisance parameter values (shear calibration bias, intrinsic alignment and source redshift bias) are chosen to be the best-fitting parameters from the analysis of \cite{des17cosmo}.  In this case, the model is given by $\vec{d} = A \vec{d}_{\rm fid}$, where $A$ is an amplitude parameter and $\vec{d}_{\rm fid}$ is the model for the correlation functions computed using the fiducial cosmological model of Table~\ref{table:prior}. The model is computed as described in Sec.~\ref{sec:modelling}.

The resultant constraints on $A$ for each redshift bin (and for the total data vector) are summarized in Table~\ref{table:bestfitA}.  We find that the measured amplitudes are consistent with $A=1$, although the first redshift bin is marginally high. We calculate the p.t.e using the $\chi^2$ of the measurement fit to the fiducial model with $A=1$ and obtain 0.14, which suggests that this deviation is not significant.
We additionally note the mild correlation between $A$ and redshift, although with our uncertainties, no conclusions could be made.

The constraint on $A$ using all redshift bins is {$A = 0.99\pm0.17$}. Furthermore, the resultant $\chi^2$ and p.t.e. values and indicate that the model is a good description of the data.  These values are shown in the rightmost columns of \Tref{table:bestfitA}. This measurement rejects the hypothesis of no lensing at a significance of {$6.8\sigma$}, and has a signal-to-noise ratio\footnote{The two values are calculated using $\sqrt{\chi^2_{\rm null}}$ and $\sqrt{\chi^2_{\rm null}-\chi^2_{\rm min}}$ respectively.} of $5.8\sigma$. {The latter value can be compared directly with results from past work: the cross-correlation measurement between Canada-France-Hawaii telescope stripe-82 survey and Atacama Cosmology Telescope obtained $4.2\sigma$ \citep{hand15}, RCSLens and $\planck$ obtained $4.2 \sigma$ \citep{hd16}, DES-SV and SPT-SZ obtained $2.9\sigma$ \citep{kirk16}, KiDS-450 and $\planck$ obtained $4.6\sigma$ \citep{hd17}. We also estimate the detection significance and signal-to-noise ratio we would have obtained with no scale cuts and find $10.8$ and $8.2\sigma$, respectively. (We note that biases due to tSZ and baryonic effects both tend to lower the cross-correlation amplitude; hence, these values are underestimates of the detection significance we would have found in the absence of these biases.)}

\subsection{Constraining shear calibration bias}
\label{sec:shearcal}

In this section and Sec.~\ref{sec:IA}, we marginalize over the cosmological parameters and nuisance parameters (shear calibration bias, intrinsic alignment and source redshift bias) simultaneously over the ranges given in Table \ref{table:prior} but combine our measurements with the \textit{Planck} baseline likelihood.\footnote{Here we use the combination of low-$\ell$ TEB and high-$\ell$ TT likelihoods.} In addition, instead of applying Gaussian priors on the shear calibration biases, we vary them over the range $[-1,1]$ and evaluate the constraining power that $w^{\gammat\kcmb}(\theta)$ has on these parameters. 

From this, we obtain $m^{2,3,4}=$[$-0.08^{+0.47}_{-0.31}$, $-0.06^{+0.20}_{-0.28}$, ${-0.14}^{+0.14}_{-0.28}$]. The data does not constrain $m^{1}$ well, which could be explained by the small overlap between the CMB lensing and the galaxy lensing kernel for this bin. These results are consistent with the constraints from cosmic shear measurements when the parameters are marginalized over in the same way: $m^{1,2,3,4}=[0.02^{+0.15}_{-0.16}, -0.04^{+ 0.09}_{-0.10}, -0.10^{+0.05}_{-0.05}, -0.05^{+0.06}_{-0.06}$], but significantly weaker than the imposed priors in \cite{des17cosmo}, which point to best-fit values of $0.012^{+0.023}_{-0.023}$ for all the bins.  These results are summarized in Table \ref{table:bestfitm}, and the posterior distributions are shown in Fig.~\ref{fig:IAconstraints}.  Our analysis demonstrates the potential of using cross-correlation measurements between galaxy lensing and CMB lensing to constrain shear calibration bias.  However, to reach the level of DES priors, the signal-to-noise of the galaxy-CMB lensing cross-correlations would have to improve by a factor of approximately 30.

\begin{table}
\centering
 \begin{tabular}{cccc}
\hline
\hline
\textsc{Sample} & $A$  & $\chi^{2}/\nu$ & p.t.e.  \\
\hline
\Tstrut
$0.20<z<0.43$  & {$1.90 \pm 0.53$} & {$2.6/3$} (0.4/4)& 0.46 (0.98)\\
$0.43<z<0.63$  & {$1.33 \pm 0.36$} & {$2.9/3$} (8.9/4)& 0.41 (0.06)\\
$0.63<z<0.90$  & {$1.04 \pm 0.22$} & {$0.7/2$} (4.3/3)& 0.69 (0.23)\\
$0.90<z<1.30$  & {$0.88 \pm 0.20$} & {$1.0/2$} (0.7/3)& 0.60 (0.87)\\[0.08cm] \hline
\Tstrut all bins & {$0.99 \pm 0.17$} & {$12.2/13$} (6.9/14)& 0.51 (0.94)\\[0.08cm]
\hline
\end{tabular}
\caption{Results of the amplitude fitting analysis described in Sec. \ref{sec:amplitude}, assuming \textit{Planck} best-fit $\Lambda$CDM cosmology. Results shown correspond to \textsc{Metacalibration} measurements with angular scale cuts applied. The numbers in enclosed in parentheses are fits for $\gammax$ to $A=0$. } 
\label{table:bestfitA}
\end{table}

\begin{table}
\centering
 \begin{tabular}{ccc}
\hline
\hline
\textsc{Sample} & $\gammat\kcmb$ & $\gamma\gamma$ \\
\hline
\Tstrut
$0.20<z<0.43$  &  $--$ & $\hspace{0.15cm}0.02^{+0.15}_{-0.16}$\\
$0.43<z<0.63$  &  $-0.08^{+0.47}_{-0.31}$ & $-0.04^{+ 0.09}_{-0.10}$\\
$0.63<z<0.90$  &  $-0.06^{+0.20}_{-0.28}$ & 
$-0.10^{+0.05}_{-0.05}$ \\
$0.90<z<1.30$  &  $-0.14^{+0.14}_{-0.28}$ & 
$-0.05^{+0.06}_{-0.06}$\\[0.08cm]
\hline
\end{tabular}
\caption{Constraints on $m^{i}$ from combining $\gammat\kcmb$ and $\gamma\gamma$  with the \textit{\planck} baseline likelihood. The constraints we obtain here are weaker than those obtained through other simulation and data based calibration methods described in \citep{Zuntz17}. }
\label{table:bestfitm}
\end{table}

\subsection{Constraining intrinsic alignment parameters}
\label{sec:IA}
\begin{figure}
\begin{center}
\includegraphics[width=1.00\linewidth]{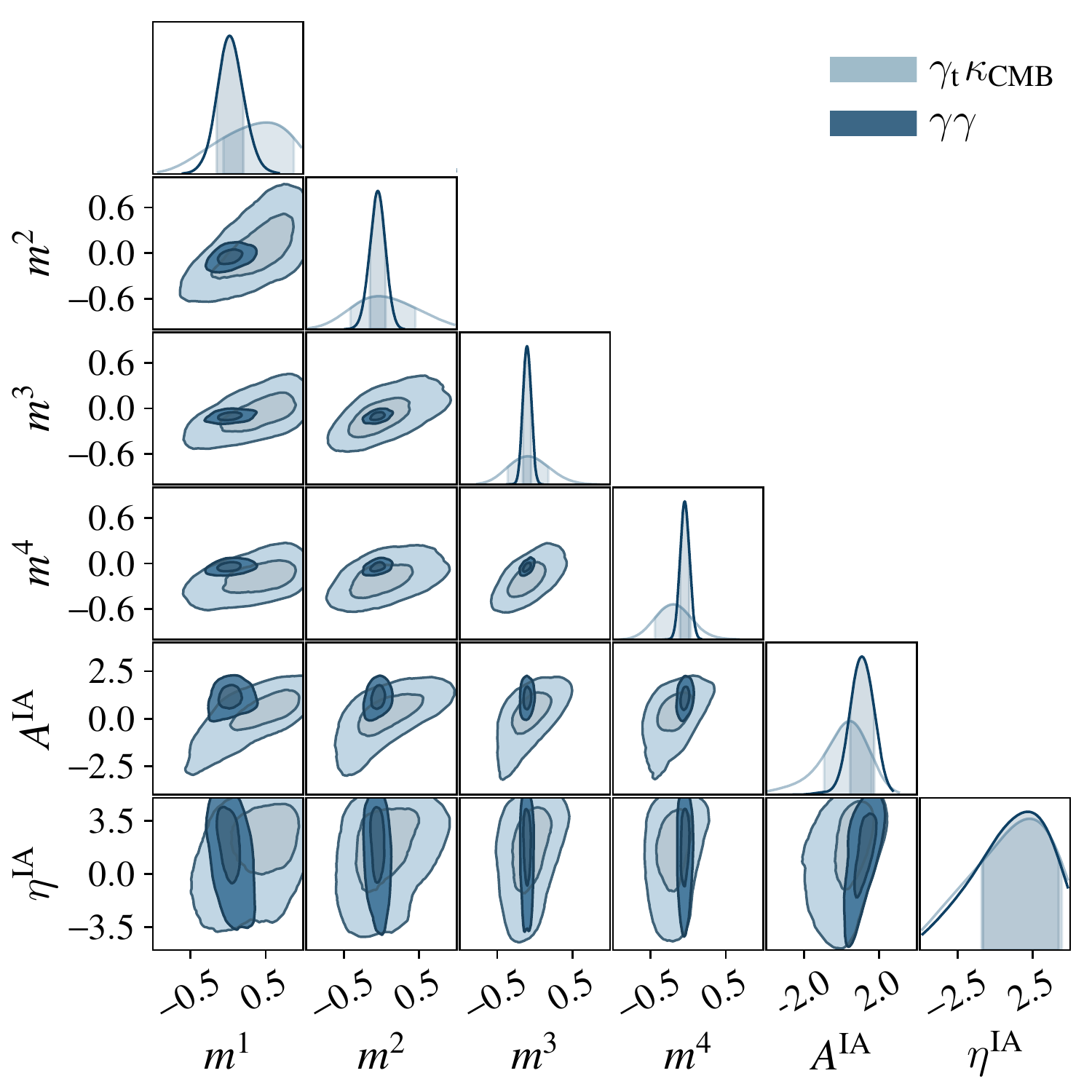}
\caption{Constraints on $m^{i}$, $A^{\rm IA}$ and $\eta^{\rm IA}$, that we marginalize over (source redshift bias parameters are also marginalized over but not shown here). The constraints that we obtain are weaker but in agreement with that from the cosmic shear measurements \citep{troxel17}.}
\label{fig:IAconstraints}
\end{center}
\end{figure}

Using the same framework as Sec.~\ref{sec:shearcal} we attempt to constrain the non-linear alignment model parameters $A^{\rm IA}$ and $\eta^{\rm IA}$. For the amplitude, we obtain $A^{\rm IA}=0.54^{+0.92}_{-1.18}$, which can be compared to  $A^{\rm IA}=1.02^{+0.64}_{-0.52}$, obtained from the DES-Y1 cosmic shear measurements. These results are in agreement with each other, although it is noted that the values are not well constrained.
Since the product of galaxy weak lensing and CMB lensing kernels span a wider redshift range compared to the galaxy weak lensing kernel alone, we might expect to obtain a better constraint on the redshift evolution parameter $\eta^{\rm IA}$ using $\gammat\kcmb$ correlations over $\gamma\gamma$. However, due to the noise level of the CMB lensing map used in this analysis, we find no significant constraint on this parameter. The results are shown in Fig. \ref{fig:IAconstraints}, and are summarized in Table \ref{table:bestfitIA}.

\begin{table}
\centering
 \begin{tabular}{cc}
\hline
\hline
\textsc{Probe} & $A^{\rm IA}$ \\
\hline
\Tstrut
$\gammat\kcmb$   &  
$0.54^{+0.92}_{-1.18}$ \\
$\gamma\gamma$  & {$1.02^{+0.64}_{-0.52}$}\\[0.08cm] 
\hline
\end{tabular}
\caption{Constraints on $A^{\rm IA}$ assuming the non-linear alignment model, when combining our $w^{\gammat\kcmb}(\theta)$ measurement and the $\planck$ baseline likelihood.}
\label{table:bestfitIA}
\end{table}
\subsection{Cosmological parameter fits}

\begin{figure}
\begin{center}
\includegraphics[width=1.0\linewidth]{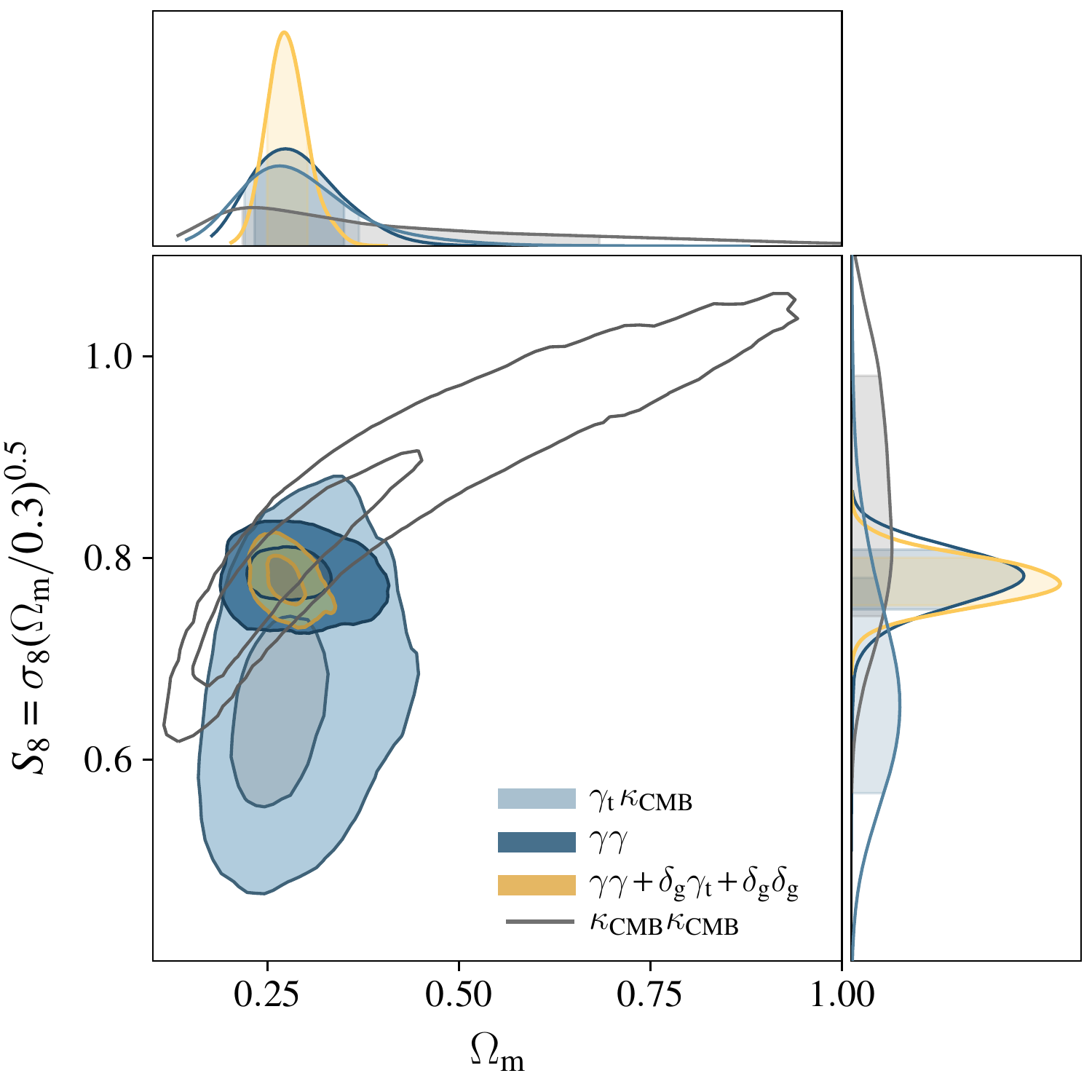}
\caption{Constraints on $\omm$ and $S_{8}$ from $\gammat\kcmb$, cosmic shear measurement from \cite{troxel17}, $3\times2$pt measurement from \cite{des17cosmo} and $\kcmb\kcmb$ measurement from \cite{Simard17}.}
\label{fig:Om_As}
\end{center}
\end{figure}

The lensing cross-correlation measurements should be sensitive to the information about the underlying dark matter distribution {and the growth of dark-matter  structure in the universe}, and hence should be sensitive to $\Omega_{\rm m}$ and  $S_{8}\equiv\sigma_{8}\sqrt{(\omm/0.3)}$. The constraints that we obtain on these parameters are shown in Fig. \ref{fig:Om_As}, and are compared with the ones obtained from the DES-Y1 cosmic shear results \citep{troxel17}, DES-Y1 joint analysis \citep{des17cosmo} and CMB lensing alone \citep{Simard17}. The comparison between our results and that of cosmic shear is interesting since we are essentially replacing one of the source planes in \cite{troxel17} with the CMB. We find that the constraints that we obtain for $w^{\gammat\kcmb}(\theta)$ are less constraining than but consistent with the cosmic shear results. The marginalized constraints on $\Omega_{\rm m}$ and $S_{8}$ are found to be {$0.261^{+0.070}_{-0.051}$} and {$0.660^{+0.085}_{-0.100}$} respectively, whereas \cite{troxel17} finds $\Omega_{\rm m}=0.260^{+0.065}_{-0.037}$ and $S_{8}=0.782^{+0.027}_{-0.027}$.

\section{Conclusions}
\label{sec:conclusion}

We have presented a measurement of the cross-correlation between galaxy lensing as measured by DES and CMB lensing as measured by SPT and \Planckc.  The galaxy lensing measurements are derived from observed distortions of the images of galaxies in approximately the redshift range of $0.2<z<1.3$; the CMB lensing measurements, on the other hand, {are inferred from distortions of the CMB temperature map induced by intervening matter along the line of sight of photons traveling from the last scattering surface.}  

The cross-correlation is detected at {$10.8\sigma$} significance including all angular bins; this is reduced to {$6.0\sigma$} after removing scales that we find to be affected by systematics such as tSZ contamination of $\kappa_{\rm CMB}$ and the effects of baryons on the matter power spectrum as described in \cite{5x2methods}. 

We perform several consistency checks on the measurements as well as tests for possible systematic errors.  These include performing null tests by cross-correlating $\kappa_{\rm CMB}$ with stellar density, dust extinction, PSF residuals and the cross-shear component, and testing our model for tSZ and CIB contamination of the $\kappa_{\rm CMB}$ map.  
{We find that of these possible systematics, the tSZ effect dominates, and we mitigate this bias by applying scale cuts to remove the angular scales that are affected the most.}

{The analytical covariance matrix that we use is tested by comparing with the jackknife covariance matrix estimated directly
from the data. The diagonal elements of these covariance matrices agree to within 25\%, which is a reasonable agreement given that the jackknife method produces a noisy estimate of the underlying covariance.}

Using the measured $w^{\gammat\kcmb}(\theta)$ correlation functions, we perform parametric fits. {Assuming a $\Lambda$CDM $\planck$ best-fit cosmology and fixing nuisance parameters to fiducial values set by DES-Y1,}  {we obtain a global best-fit amplitude of} {$A=0.99\pm0.17$} {which is} consistent with expectations from the $\Lambda$CDM cosmological model ($A=1$). 

Next, we combine our measurement with the $\planck$ baseline likelihood, and vary the nuisance parameters and attempt to constrain them. For the shear calibration bias parameters  we obtain the constraints $m^{2,3,4}=$[ $-0.08^{+0.47}_{-0.31}$, $-0.06^{+0.20}_{-0.28}$, ${-0.14}^{+0.14}_{-0.28}$], while $m^{1}$ is not constrained well. These constraints are less stringent than the DES-Y1 priors derived from data and simulations, it is anticipated that the $\gammat\kcmb$ correlation will be able to constrain shear calibration bias to better precision than these methods \citep{Schaan17} for future surveys such as CMB-S4 \citep{S4} and LSST \citep{LSST}. 

For the amplitude of IA, we obtain the constraint $A^{\rm IA}=0.54^{+0.92}_{-1.18}$, which is in agreement with what is obtained from DES-Y1 cosmic shear measurements. However, the redshift evolution parameter $\eta^{\rm IA}$ is not constrained well using $w^{\gammat\kcmb}(\theta)$ measurement alone. 

When we marginalize over the nuisance parameters using the DES-Y1 priors listed in Table \ref{table:prior}, we obtain constraints on cosmological parameters that are consistent with recent results from \cite{Zuntz17}: {$\Omega_{\rm m}=0.261^{+0.070}_{-0.051}$} and {$S_{8}\equiv\sigma_{8}\sqrt{\Omega_{\rm m}/0.3} =0.660^{+0.085}_{-0.100}$}. {While the constraining power of $\gammat\kcmb$ is relatively weak, we obtain independent constraints on $\Omega_{\rm m}$ and $S_{8}$, which will help break degeneracies in parameter space when all the probes are combined. 

Future data from the full DES survey and SPT-3G \citep{Benson2014} should provide significant reduction in measurement uncertainties on the $w^{\gammat\kcmb}(\theta)$ correlation function. Moreover, tSZ contamination of the temperature-based CMB lensing map necessitates removal of certain angular scales, which reduces the signal-to-noise of the measurements significantly. For SPT-3G, the CMB lensing map will be reconstructed using polarisation data, which will have minimal tSZ  contamination.\footnote{The lensing potential can also be reconstructed using combinations of temperature maps such that the lensing map is less sensitive to biases; see e.g. \cite{Madhavacheril2018}.}  {With these potential improvements, the $\gammat\kcmb$ cross-correlation is a promising probe from which will be used to extract constraints independent of those from galaxy shear or CMB measurements alone.} 

\section*{Acknowledgements}

YO acknowledges funding from the Natural Sciences and Engineering Research Council of Canada, Canadian Institute for Advanced Research, Canada Research Chairs program and support from the Kavli Foundation. EB is partially supported by the US Department of Energy grant DE-SC0007901. CC was supported in part by the Kavli Institute for Cosmological Physics at the University of Chicago through grant NSF PHY-1125897 and an endowment from Kavli Foundation and its founder Fred Kavli.

Computations were made on the supercomputer Guillimin from McGill University, managed by Calcul Qu\'{e}bec and Compute Canada. The operation of this supercomputer is funded by the Canada Foundation for Innovation (CFI), the minist\`{e}re de l'\'{E}conomie, de la science et de l'innovation du Qu\'{e}bec (MESI) and the Fonds de recherche du Qu\'{e}bec - Nature et technologies (FRQ-NT).

This research is part of the Blue Waters sustained-petascale computing project, which is supported by the National Science Foundation (awards OCI-0725070 and ACI-1238993) and the state of Illinois. Blue Waters is a joint effort of the University of Illinois at Urbana-Champaign and its National Center for Supercomputing Applications.

This research used resources of the National Energy Research Scientific Computing Center (NERSC), a DOE Office of Science User Facility supported by the Office of Science of the U.S. Department of Energy under Contract No. DE-AC$02$-$05$CH$11231$.

The South Pole Telescope program is supported by the National Science Foundation through grant PLR-1248097. Partial support is also provided by the NSF Physics Frontier Center grant PHY-0114422 to the Kavli Institute of Cosmological Physics at the University of Chicago, the Kavli Foundation, and the Gordon and Betty Moore Foundation through Grant GBMF\#947 to the University of Chicago. The McGill authors acknowledge funding from the Natural Sciences and Engineering Research Council of Canada, Canadian Institute for Advanced Research, and Canada Research Chairs program.
CR acknowledges support from a Australian Research Council Future Fellowship (FT150100074)
BB is supported by the Fermi Research Alliance, LLC under Contract No. De-AC02-07CH11359 with the United States Department of Energy.

Argonne National Laboratory’s work was supported under U.S. Department of Energy contract DE-AC02-06CH11357.
Funding for the DES Projects has been provided by the U.S. Department of Energy, the U.S. National Science Foundation, the Ministry of Science and Education of Spain, 
the Science and Technology Facilities Council of the United Kingdom, the Higher Education Funding Council for England, the National Center for Supercomputing 
Applications at the University of Illinois at Urbana-Champaign, the Kavli Institute of Cosmological Physics at the University of Chicago, 
the Center for Cosmology and Astro-Particle Physics at the Ohio State University,
the Mitchell Institute for Fundamental Physics and Astronomy at Texas A\&M University, Financiadora de Estudos e Projetos, 
Funda{\c c}{\~a}o Carlos Chagas Filho de Amparo {\`a} Pesquisa do Estado do Rio de Janeiro, Conselho Nacional de Desenvolvimento Cient{\'i}fico e Tecnol{\'o}gico and 
the Minist{\'e}rio da Ci{\^e}ncia, Tecnologia e Inova{\c c}{\~a}o, the Deutsche Forschungsgemeinschaft and the Collaborating Institutions in the Dark Energy Survey. 

The Collaborating Institutions are Argonne National Laboratory, the University of California at Santa Cruz, the University of Cambridge, Centro de Investigaciones Energ{\'e}ticas, 
Medioambientales y Tecnol{\'o}gicas-Madrid, the University of Chicago, University College London, the DES-Brazil Consortium, the University of Edinburgh, 
the Eidgen{\"o}ssische Technische Hochschule (ETH) Z{\"u}rich, 
Fermi National Accelerator Laboratory, the University of Illinois at Urbana-Champaign, the Institut de Ci{\`e}ncies de l'Espai (IEEC/CSIC), 
the Institut de F{\'i}sica d'Altes Energies, Lawrence Berkeley National Laboratory, the Ludwig-Maximilians Universit{\"a}t M{\"u}nchen and the associated Excellence Cluster Universe, 
the University of Michigan, the National Optical Astronomy Observatory, the University of Nottingham, The Ohio State University, the University of Pennsylvania, the University of Portsmouth, 
SLAC National Accelerator Laboratory, Stanford University, the University of Sussex, Texas A\&M University, and the OzDES Membership Consortium.

Based in part on observations at Cerro Tololo Inter-American Observatory, National Optical Astronomy Observatory, which is operated by the Association of 
Universities for Research in Astronomy (AURA) under a cooperative agreement with the National Science Foundation.

The DES data management system is supported by the National Science Foundation under Grant Numbers AST-1138766 and AST-1536171.
The DES participants from Spanish institutions are partially supported by MINECO under grants AYA2015-71825, ESP2015-66861, FPA2015-68048, SEV-2016-0588, SEV-2016-0597, and MDM-2015-0509, 
some of which include ERDF funds from the European Union. IFAE is partially funded by the CERCA program of the Generalitat de Catalunya.
Research leading to these results has received funding from the European Research
Council under the European Union's Seventh Framework Program (FP7/2007-2013) including ERC grant agreements 240672, 291329, and 306478.
We  acknowledge support from the Australian Research Council Centre of Excellence for All-sky Astrophysics (CAASTRO), through project number CE110001020.

This manuscript has been authored by Fermi Research Alliance, LLC under Contract No. DE-AC02-07CH11359 with the U.S. Department of Energy, Office of Science, Office of High Energy Physics. The United States Government retains and the publisher, by accepting the article for publication, acknowledges that the United States Government retains a non-exclusive, paid-up, irrevocable, world-wide license to publish or reproduce the published form of this manuscript, or allow others to do so, for United States Government purposes.

We acknowledge the use of many python packages: \textsc{Astropy}, a community-developed core Python package for Astronomy \citep{astropy18}, \textsc{CAMB} \citep{Lewis00,Howlett12}, \textsc{Chain consumer}\footnote{\url{https://samreay.github.io/ChainConsumer}}, \textsc{CosmoSIS}\footnote{\url{https://bitbucket.org/joezuntz/cosmosis}},
\textsc{HEALPix} \citep{gorski05}, \textsc{IPython} \citep{ipython07}, \textsc{Matplotlib} \citep{hunter07}, \textsc{NumPy} \& \textsc{SciPy} \citep{scipy,numpy}, \textsc{Quicklens}\footnote{\url{https://github.com/dhanson/quicklens}} and \textsc{TreeCorr} \citep{jarvis15}.

\bibliography{ref} 



\appendix

\section{Flask simulations}
\label{sec:flask}
In this work, we make use of the publicly available code \flask\ \citep{Xavier:2016}, to generate correlated maps between shear and CMB lensing.  We use \flask\ to generate 120 full-sky log-normal realizations of the density field and four galaxy shear maps corresponding to the four redshift bins we use for the data. Additionally, we generate a convergence map at $z=1089$, and treat this as a noiseless CMB convergence map.  The galaxy shear catalogs are generated using galaxy number densities and shape noise measured from data, and Gaussian noise realizations generated from the noise power spectrum of the CMB convergence maps are added to the noiseless convergence map to produce data-like catalogs and maps. For each full sky simulation, we extract out ten sub-catalogs by applying the DES-Y1 angular mask, resulting in 1200 synthetic galaxy shear catalogs and CMB convergence maps that have noise properties matched to the real data.

\section{Validation of jackknife covariance estimate}
\label{sec:jackknife_validation}
To test whether the jackknife covariance estimate provides a reliable estimate of the true covariance over the scales considered, we make use of $\flask$ simulation realizations. For each of the simulated catalogues, we measure $w^{\gammat\kcmb}(\theta)$ using the same procedure as applied to the real data.  We then compute the covariance matrix directly across the 1200 simulated catalogs, which provides a low-noise estimate of the covariance of $w^{\gammat\kcmb}(\theta)$ in the \flask\ simulations. (which we call ``true" $\flask$ covariance).  From the simulated catalogue, we also compute the jackknife estimate of the covariance and compare this with the true $\flask$ covariance. We find that these are consistent with each other to within 25\%.


\label{lastpage}

\end{document}

%% file: journaldef.tex
\def\aj{AJ}%
\def\apj{ApJ}%
\def\apjs{ApJS}%
\def\ao{Appl.~Opt.}%
\def\aap{A\&A}%
\def\mnras{MNRAS}%
\def\prd{Phys.~Rev.~D}%
\def\jcap{JCAP}%
\def\physrep{Phys.~Rep.}%
\def\procspie{Proc.~SPIE}%

%% file: notations.tex
\newcommand{\omm}{\Omega_{\rm m}}

\newcommand{\omb}{\Omega_{\rm b}}

\newcommand{\w}{w_0}

\newcommand{\sqdeg}{{\rm deg}^{2}}

\def\3x2pt{3$\times$2pt}
\def\5x2pt{5$\times$2pt}
\def\6x2pt{6$\times$2pt}

\newcommand{\planck}{\textit{Planck}}

\newcommand{\kcmb}{\kappa_{\rm CMB}}
\newcommand{\ktsz}{\kappa_{\rm tSZ}}

\newcommand{\kgal}{\kappa_{\rm s}}

\newcommand{\gammat}{\gamma_{\rm t}}
\newcommand{\gammax}{\gamma_{\times}}

\newcommand{\imshape}{\textsc{Im3shape}}

\newcommand{\treecorr}{\textsc{TreeCorr}}

\newcommand{\metacal}{\textsc{Metacalibration}}
\newcommand{\flask}{\textsc{Flask}}

\newcommand{\redmagic}{\textsc{redMaGiC}}
\newcommand{\redmapper}{\textsc{redMaPPer}}

%% file: authorlist.tex

\author{Y.~Omori}
\affiliation{Kavli Institute for Particle Astrophysics \& Cosmology, P. O. Box 2450, Stanford University, Stanford, CA 94305, USA}
\affiliation{Department of Physics, Stanford University, 382 Via Pueblo Mall, Stanford, CA 94305, USA}
\affiliation{Department of Physics and McGill Space Institute, McGill University, Montreal, Quebec H3A 2T8, Canada}
\author{E.~J.~Baxter}
\affiliation{Department of Physics and Astronomy, University of Pennsylvania, Philadelphia, PA 19104, USA}
\author{C.~Chang}
\affiliation{Kavli Institute for Cosmological Physics, University of Chicago, Chicago, IL 60637, USA}
\author{D.~Kirk}
\affiliation{Department of Physics \& Astronomy, University College London, Gower Street, London, WC1E 6BT, UK}
\author{A.~Alarcon}
\affiliation{Institut d'Estudis Espacials de Catalunya (IEEC), 08193 Barcelona, Spain}
\affiliation{Institute of Space Sciences (ICE, CSIC),  Campus UAB, Carrer de Can Magrans, s/n,  08193 Barcelona, Spain}
\author{G.~M.~Bernstein}
\affiliation{Department of Physics and Astronomy, University of Pennsylvania, Philadelphia, PA 19104, USA}
\author{L.~E.~Bleem}
\affiliation{High Energy Physics Division, Argonne National Laboratory, Argonne, IL, USA 60439}
\affiliation{Kavli Institute for Cosmological Physics, University of Chicago, Chicago, IL 60637, USA}
\author{R.~Cawthon}
\affiliation{Kavli Institute for Cosmological Physics, University of Chicago, Chicago, IL 60637, USA}
\author{A.~Choi}
\affiliation{Center for Cosmology and Astro-Particle Physics, The Ohio State University, Columbus, OH 43210, USA}
\author{R.~Chown}
\affiliation{2Department of Physics and Astronomy, McMaster University, 1280 Main St. W., Hamilton, ON L8S 4L8, Canada}
\affiliation{Department of Physics and McGill Space Institute, McGill University, Montreal, Quebec H3A 2T8, Canada}
% ...

\collaboration{DES \& SPT Collaborations}